\documentclass[aps,prb, superscriptaddress ,twocolumn,10pt,final,times]{revtex4}

\usepackage{graphicx}
\usepackage{color}
\usepackage{amssymb}
\usepackage{amsmath}
\usepackage{wasysym} 
\usepackage{hyperref}
\usepackage{times}
\begin{document}

\title{{\fontfamily{ptm}\selectfont \Large Entanglement of spin waves among four quantum memories }}

\author{\fontfamily{ptm}\selectfont K. S. Choi}
\affiliation{\fontfamily{ptm}\selectfont Norman Bridge Laboratory of Physics 12-33, California Institute of Technology, Pasadena, California 91125, USA}
\author{\fontfamily{ptm}\selectfont A. Goban}
\affiliation{\fontfamily{ptm}\selectfont Norman Bridge Laboratory of Physics 12-33, California Institute of Technology, Pasadena, California 91125, USA}
\author{\fontfamily{ptm}\selectfont S. B. Papp$^{\dagger}$\footnotetext{\small $^{\dagger}$ Present address : National Institute of Standards and Technology, Boulder, Colorado 80305, USA} }
\affiliation{\fontfamily{ptm}\selectfont Norman Bridge Laboratory of Physics 12-33, California Institute of Technology, Pasadena, California 91125, USA}
\author{\fontfamily{ptm}\selectfont  S. J. van Enk}
\affiliation{\fontfamily{ptm}\selectfont Department of Physics, University of Oregon, Eugene, Oregon 97403, USA}
\author{\fontfamily{ptm}\selectfont H. J. Kimble}
\affiliation{\fontfamily{ptm}\selectfont Norman Bridge Laboratory of Physics 12-33, California Institute of Technology, Pasadena, California 91125, USA}
\date{\today}

\maketitle

\fontfamily{ptm}\selectfont
\noindent\textbf{\fontfamily{ptm}\selectfont Quantum networks are composed of quantum nodes that interact coherently by way of quantum channels and open a broad frontier of scientific opportunities \cite{kimble08}. For example, a quantum network can serve as a `web' for connecting quantum processors for computation \cite{preskill97,nielson00} and communication \cite{duan01}, as well as a `simulator' for enabling investigations of quantum critical phenomena arising from interactions among the nodes mediated by the channels \cite{acin07,illuminati06}. The physical realization of quantum networks generically requires dynamical systems capable of generating and storing entangled states among multiple quantum memories, and of efficiently transferring stored entanglement into quantum channels for distribution across the network. While such capabilities have been demonstrated for diverse bipartite systems (i.e., $N=2$ quantum systems) \cite{chou05, wilk07, moehring07,simon07,choi08, jost09}, entangled states with $N > 2$ have heretofore not been achieved for quantum interconnects that coherently `clock' multipartite entanglement stored in quantum memories to quantum channels. Here, we demonstrate high-fidelity measurement-induced entanglement stored in four atomic memories; user-controlled, coherent transfer of atomic entanglement to four photonic quantum channels; and the characterization of the full quadripartite entanglement by way of quantum uncertainty relations \cite{hofmann03,papp09,lougovski09}. Our work thereby provides an important tool for the distribution of multipartite entanglement across quantum networks. Moreover, we show how our entanglement verification method, originally developed for infinite-dimensional bosonic systems, also applies to finite-dimensional quantum spin-$1/2$ systems, where our results may find application for investigations of entanglement order in condensed matter systems at thermal equilibrium\cite{amico08,guhne09}. With regard to quantum measurements, the multipartite entangled state stored in the quantum memories can be applied for sensing an atomic phase shift beyond the limit for any unentangled state.}

\vspace{0.2 cm}

Diverse applications in quantum information science require coherent control of the generation, storage, and transfer of entanglement among spatially separated physical systems \cite{kimble08,preskill97,nielson00, duan01,acin07}. Despite its inherently multipartite nature, entanglement has been studied primarily for bipartite systems \cite{nielson00,clauser78,aspect81}, where remarkable progress has been made in harnessing physical processes to generate `push-button' and `heralded' entanglement\cite{chou05, wilk07, moehring07,simon07}, as well as to map entangled states to and from atoms, photons, and phonons\cite{choi08, jost09}. In this endeavor, well-established methods for characterizing bipartite entanglement of discrete and continuous quantum variables have been essential\cite{nielson00,guhne09,clauser78,aspect81,horodecki09,vanenk07}.

\begin{figure*}[t!]
\includegraphics[width=1.85\columnwidth]{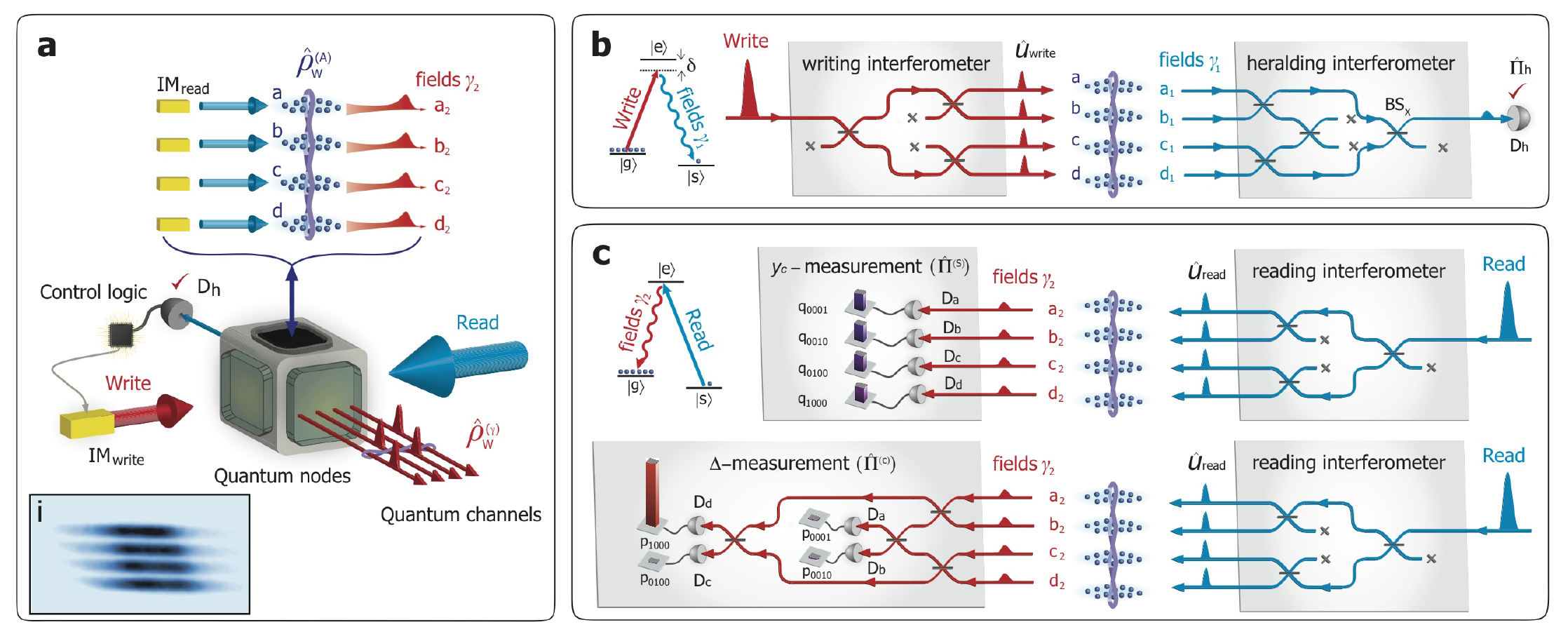}
\caption{\fontfamily{ptm}\selectfont \small\textbf{Overview of the experiment.}
\textbf{a,} Quantum memories and interfaces for multipartite quantum networks. Entangled state $\hat{\rho}^{(A)}_{W}$ for four atomic ensembles $\epsilon=\{a,b,c,d\}$ is heralded by a photoelectric detection event at detector $D_{h}$ derived from quantum interference of four fields $\gamma_1=\{a_{1},b_{1},c_{1},d_{1}\}$. After a user-defined delay $\tau$, read lasers are applied to the individual ensembles to coherently transform the atomic entangled state $\hat{\rho}^{(A)}_{W}$ into quadripartite entangled beams of light $\hat{\rho}^{(\gamma)}_{W}$, with each beam propagating through quantum channels $\gamma_2=\{a_{2},b_{2},c_{2},d_{2}\}$. Intensity modulators $\text{IM}_{\text{write,read}}$ control the intensities of the writing and reading beams. Inset \textbf{i}, a fluorescence image of the laser-cooled atomic samples $\{a,b,c,d\}$ (Appendix). 
\textbf{b,} Entanglement generation. A weak write laser is split into four components to excite atomic ensembles $\{a,b,c,d\}$ via parametric interactions $\hat{U}_{\text{write}}$. A single Raman scattered photon for four fields $\gamma_1$ is emitted by the ensembles and detected by a projective measurement $\hat{\Pi}_{h}$ at detector $D_h$, which signals the creation of an entangled state $\hat{\rho}^{(A)}_{W}$ for ensembles $\{a,b,c,d\}$ (Eq. \ref{atomWstate}).
\textbf{c,} Quantum-state exchange and entanglement verification. A strong read pulse is sent into the four atomic ensembles, and the atomic state $\hat{\rho}^{(A)}_{W}$ is mapped to an entangled state of light $\hat{\rho}^{(\gamma)}_{W}$ for four fields $\gamma_2$ (Eq. \ref{photonWstate}) via quantum-state transfers, $\hat{U}_{\text{read}}$. This entangled field state then propagates to the entanglement verification ports. (i) Upper panel for $y_c$-measurement $-$ The quantum statistics $\{q_{i,j,k,l}\}$ of $\hat{\rho}^{(\gamma)}_{W}$ are measured with projectors $\{\hat{\Pi}^{(s)}_{i}\}$ for the modes at detectors $D_{a,b,c,d}$. (ii) Lower panel for $\Delta$-measurement $-$ By rerouting the relevant fibre-optical connections, we access mutual coherences for $\hat{\rho}^{(A)}_{W}$ with projectors $\{\hat{\Pi}^{(c)}_{i}\}$ from detection statistics $p_{ijkl}$ at $D_{a,b,c,d}$. States $|g\rangle$, $|s\rangle$ are hyperfine ground states $F=4$, $F=3$ of $6S_{1/2}$ in atomic Caesium; state $|e\rangle$ is the hyperfine level $F^{\prime}=4$ of the electronic excited state $6P_{3/2}$.}
\label{fig1}
\end{figure*}

For multipartite systems, the `size' of a physical state characterized by its density matrix $\hat{\rho}_{N}$ grows exponentially with the number of subsystems $N$ and makes the entangled states exceedingly difficult to represent with classical information. Importantly, this complexity for $\hat{\rho}_{N}$ increases the potential utility of multipartite entanglement in quantum information science, including for quantum algorithms \cite{nielson00,preskill97} and simulation\cite{lloyd96}. Redundant encoding of quantum information into multipartite entangled states enables quantum error correction and fault tolerant quantum computation\cite{preskill97,nielson00}.  The intricate long-range quantum correlations of quantum many-body systems are intertwined with multipartite entanglement in a fashion that is a subject of active investigation\cite{amico08,guhne09}. In addition, mobilizing multipartite entanglement across quantum networks could lead to novel quantum phase transitions for the network, as for the percolation of entanglement\cite{acin07}. 

Counterposed to these opportunities, the complex structure of multipartite entanglement presents serious challenges both for its formal characterization and physical realization. Theoretically, there is no unique measure of $M$-partite entanglement for $M>2$ components of the $N$-partite $\hat{\rho}_{N}$, nor is there a complete classification of the types of entanglement\cite{horodecki09}. Apart from the technical difficulty of preparing $\hat{\rho}_{N}$ for a large system, there is only a nascent understanding of robust means for entanglement verification and characterization that incorporate experimental imperfections and finite measurement resources. Indeed, there are relatively few examples of laboratory systems that have successfully generated multipartite entanglement. Most works have considered the entanglement for spin systems, notably trapped ions\cite{leibfried05a,haffner05a,barreiro10}, which are applicable to the matter nodes of a quantum network. But the methodologies employed for entanglement characterization are problematic for the infinite dimensional bosonic systems of the quantum channels (e.g., quadrature\cite{aoki03} and number-state\cite{papp09} entanglement for optical modes). A relevant advance for photonic systems has been \textit{a posteriori} entanglement generated from parametric downconversion\cite{gao08}. However, in this case the post-dicted state $\hat{\sigma}_{N}$ is obtained by a destructive local filter, is only a small component of a larger physical state $\hat{{\rho}}_{N}$, and is not available for subsequent utilization\cite{vanenk07}.

In addition to the generation and characterization of multipartite entanglement, an important capability for quantum networks is the development of quantum interfaces capable of generating, storing, and dynamically allocating the entanglement of matter nodes into photonic (or phononic) channels. As illustrated in Fig. \ref{fig1}a, we introduce here such a quantum interface for quadripartite entangled states based upon coherent, collective emission from matter to light. We present a systematic study of the generation and storage of novel quadripartite entangled states of spin-waves in a set of four atomic memories, which could be located within one node or distributed at multiple nodes, as well as of the coherent transfer of the entangled components of the material state into individual photonic channels. We measure transitions from $M=4$ to $3$ to $2$-partite entanglement and to a fully separable state as a function of controlled spin-wave statistics of the atomic memories. We also explore the temporal decoherence of the atomic entanglement and observe a dynamic evolution of multipartite entanglement into various subsets of the full quadripartite states in a dissipative environment, from fully quadripartite entangled to unentangled.

As illustrated in Fig. \ref{fig1}, our experiment proceeds in four steps (see Appendix). First, in step (1) an entangled state $\hat{\rho}^{(A)}_{W}$ of four atomic ensembles is generated by quantum interference in a quantum measurement\cite{duan01,chou05} (Fig. \ref{fig1}b). Given a photoelectric detection event at $D_h$, the conditional atomic state is ideally a quadripartite entangled state $\hat{\rho}^{(A)}_{W} = | W \rangle_{_{A}}\langle W |$ with
\begin{equation}
\begin{array}{rcl}
\label{atomWstate}
|W\rangle_{_{A}}=\frac{1}{2}[(
|\overline{s}_{a}, \overline{g}_{b}, \overline{g}_{c}, \overline{g}_{d}\rangle+e^{i\phi_{1}}|\overline{g}_{a}, \overline{s}_{b}, \overline{g}_{c}, \overline{g}_{d}\rangle) +\\\\
e^{i\phi_{2}}(|\overline{g}_{a}, \overline{g}_{b}, \overline{s}_{c}, \overline{g}_{d}\rangle+e^{i\phi_{3}}|\overline{g}_{a}, \overline{g}_{b}, \overline{g}_{c}, \overline{s}_{d}\rangle)],&
\end{array}
\end{equation} 
whose single quantum spin-wave $|\overline{s}_{\epsilon}\rangle$ is coherently shared among four ensembles $\epsilon=\{a,b,c,d \}$. These entangled states are known as $W$-states, whose component states are the atomic ground states $|\overline{g}_{\epsilon}\rangle=|g \cdots g\rangle_{\epsilon}$ and single collective excitations $|\overline{s}_{\epsilon}\rangle=\frac{1}{\sqrt{N_{A,\epsilon}}}\sum_{i=1}^{N_{A,\epsilon}} |g \cdots s_{i} \cdots g \rangle_{\epsilon}$, where $N_{A,\epsilon}$ is the number of atoms in ensemble $\epsilon $.

After the heralding event at $D_h$, step (2) consists of storage of $\hat{\rho}^{(A)}_{W}$ in the ensembles without optical illumination for a user-controlled time $\tau$. At the end of this interval, step (3) is initiated with read beams applied to the individual ensembles to coherently transfer the respective entangled components of $\hat{\rho}^{(A)}_{W}$ into a quadripartite entangled state of light $\hat{\rho}^{(\gamma)}_{W}=|W\rangle_{\gamma}\langle W|$ via cooperative emissions\cite{duan01} (Fig. \ref{fig1}c), where 
\begin{equation}
\label{photonWstate}
|W\rangle_{\gamma}=\frac{1}{2}[(
|1000\rangle+e^{i\phi^{\prime}_{1}}|0100\rangle)+e^{i\phi^{\prime}_{2}}(|0010\rangle+e^{i\phi^{\prime}_{3}}|0001\rangle)].
\end{equation}
This photonic state is a mode-entangled $W$-state\cite{papp09,lougovski09}, analogous to Eq. \ref{atomWstate}, which shares a single delocalized photon among four optical modes $\gamma_2=\{a_{2},b_{2},c_{2},d_{2}\}$. Because the mappings of quantum states from the collective atomic modes of the ensembles to individual field modes are local operations, the presence of quadripartite atomic entanglement for $\hat{\rho}^{(A)}_{W}(\tau)$ can be unambiguously determined by accessing the entanglement degree of the photonic state $\hat{\rho}^{(\gamma)}_{W}(\tau)$ via the state transfers $\hat{U}_{\text{read}}$ at time $\tau$.

Finally, in step (4) we verify and characterize the heralded entanglement for $\hat{\rho}^{(\gamma)}_{W}$ from complementary measurements of photon statistics and coherence \cite{papp09,lougovski09} (Fig. \ref{fig1}c). In particular, we consider the reduced density matrix $\hat{\rho}_{r}=p_0 \hat{\rho}_0+p_1 \hat{\rho}_1+p_{\geq 2} \hat{\rho}_{\geq 2}$ containing up to one photon per mode, which leads to a lower bound for the entanglement of the actual physical state $\hat{\rho}^{(\gamma)}_{W}$ in an infinite dimensional bosonic space. In turn, the truncation provides a lower bound inference for the atomic entanglement in $\hat{\rho}^{(A)}_{W}$. Here, $\{p_{0}, p_{1}, p_{\geq 2}\}$ are the probabilities for the $0$ and $1$-photon subspaces $\hat{\rho}_{0,1}$ and the higher-order subspace $\hat{\rho}_{\geq 2}$, respectively. As illustrated in the upper panel of Fig. \ref{fig1}c, these quantities are evaluated from the probabilities $q_{ijkl}$ for $i,j,k,l\in\{0,1\}$ photons to occupy the respective optical modes $\gamma_2=\{a_{2},b_{2},c_{2},d_{2}\}$ at the output faces of the ensembles via photoelectric detections $\{ \hat{\Pi}^{(s)}_{i}\}$ at detectors $D_{a,b,c,d}$. The photon probabilities $\{p_{0}, p_{1}, p_{\geq 2}\}$ are combined into a normalized measure of the degree of statistical contamination from multiple excitations $\hat{\rho}_{\geq 2}$ for $\hat{\rho}^{(\gamma)}_{W}$, namely $y_c\equiv\frac{8}{3}\frac{p_{\geq 2}p_0}{p_1^2}$, with $y_c=0$ for a single excitation (i.e., $ p_{\geq 2}=0$) to $y_c=1$ for balanced coherent states\cite{lougovski09}. Operationally, we control $y_c$ (thereby, the spin-wave statistics for $\hat{\rho}^{(A)}_{W}$ and the photon statistics for $\hat{\rho}^{(\gamma)}_{W}$) by way of the intensity of the write beam.

We must also quantify the mutual coherences $d_{\alpha\beta}$ for the optical modes $\alpha,\beta\in \{a_{2},b_{2},c_{2},d_{2}\}$ of $\hat{\rho}^{(\gamma)}_{W}$. This is accomplished by measuring the photon probabilities $\{p_{1000},p_{0100},p_{0010},p_{0001}\}$ at the outputs of the verification ($v$) interferometer shown in the lower panel of Fig. \ref{fig1}c, from which we determine the sum uncertainty $\Delta\equiv\sum_{i=1}^{N=4}\langle (\hat{\Pi}^{(c)}_{i})^2-\langle\hat{\Pi}^{(c)}_{i}\rangle^{2}\rangle$ for the collective variables $\{\hat{\Pi}^{(c)}_{i}\}=\{|W_i\rangle_v\langle W_i|\}$. Here, $\{|W_i\rangle_v\}$ is a set of four orthonormal $W$-states, with phases $\{\beta_1,\beta_2,\beta_3\}_v$ for the set selected by way of the actively stabilized paths in the verification interferometer. For appropriate choices of $\{\beta_1,\beta_2,\beta_3\}_v$, photodetections at $D_i$ act as projective measurements $\hat{\Pi}^{(c)}_{i}$ of the input $\hat{\rho}_r$ onto $|W_i\rangle_v$. From our measurement of $\Delta$, we deduce an effective visibility $V_{\text{eff}}=4\overline{d}$ for the interference of any two of the four modes $\{a_{2},b_{2},c_{2},d_{2}\}$, where $\overline{d}=\text{avg}(d_{\alpha\beta})$ with $0\leq\overline{d}\leq\frac{1}{4}$. Hence, for the ideal $W$-state in Eq. \ref{photonWstate}, we have $d_{\alpha\beta}=1/4$, so that $\Delta=0$ and $V_{\text{eff}}=1$, with associated photon probabilities $p_{1000}=1$ and $p_{0100}=p_{0010}=p_{0001}=0$. 
By comparison, we obtain $p_{1000}\simeq 0.97\pm 0.01$,  $p_{0100}\simeq p_{0010}\simeq p_{0001}\simeq 0.010\pm 0.003$ in the experiment for $y_c=0.04\pm 0.01$, as shown in the bar plots of the lower panel of Fig. \ref{fig1}c. In contrast, a mixed state analogous to Eq. \ref{photonWstate} but with no phase coherence $d_{\alpha\beta}=0$ would result in balanced photon probabilities (i.e., $p_{1000}=p_{0100}=p_{0010}=p_{0001}=1/4$) at the outputs of the interferometer, and thereby yield $\Delta=0.75$ (i.e., $V_{\text{eff}}=0$). More generally, the degree of imbalance for $p_{ijkl}$ can be expressed in terms of $d_{\alpha\beta}$, and leads to the sum uncertainty $\Delta\lesssim\frac{3}{4}(1-16\overline{d}^2)$.

The pair $\{\Delta,y_c\}$ thereby defines the parameter space for the multipartite entanglement employed in our experiment
, with the entanglement parameters $\{\Delta,y_c\}$ serving as a nonlocal, nonlinear entanglement witness \cite{lougovski09}. Specifically, for a given value of $y_c$, we numerically determine the bound $\Delta^{(M-1)}_b$ for the minimal uncertainty possible for all states containing at most $(M-1)$-mode entanglement and their mixtures. For our quadripartite states $N=4$, we derive the respective uncertainty bounds $\{\Delta^{(3)}_b,\Delta^{(2)}_b,\Delta^{(1)}_b\}$ for tripartite entangled, bipartite entangled and fully separable states, as functions of $y_c$. A measurement of quantum statistics $y_c$ and the associated coherence $\Delta$ with $\Delta < \Delta^{(1,2,3)}_b$ thereby manifestly confirms the presence of genuine $M=4$ partite entanglement, with refs. \cite{papp09,lougovski09} providing further details. 

We emphasize that our criterion of `genuine' $M$-partite entanglement based upon $\{\Delta,y_c\}$ takes the most stringent form of non-separability \cite{horodecki09} and excludes all weaker forms of multipartite entanglement, including those which cannot be merely separated into two groups along any splittings of $M$ (ref. \cite{lougovski09}). The genuine $M$-partite entangled state created from our experiment can only be represented as mixtures of pure states that \textit{all} possess $M$-partite entanglement (similar to the notion of \textit{k-producibility} in multipartite spin models\cite{amico08,guhne09}). Furthermore, our verification protocol is capable of unambiguously distinguishing genuine $M$ and $(M-1)$-partite entangled states for any $M\leq N$ by observing the uncertainty $\Delta$ below $\Delta^{(M-1)}_b$. We take caution that our entanglement verification protocol cannot be applied for verifying the \textit{absence} of entanglement for the physical state $\hat{\rho}^{(\gamma)}_{W}$ of infinite dimensions\cite{eisert02}, as our analysis assumes a finite-dimensional truncated $\hat{\rho}_r$ of $\hat{\rho}^{(\gamma)}_{W}$ leading to lower bound entanglement \cite{lougovski09}.

\begin{figure}[tbph!]
\includegraphics[width= 0.7\columnwidth]{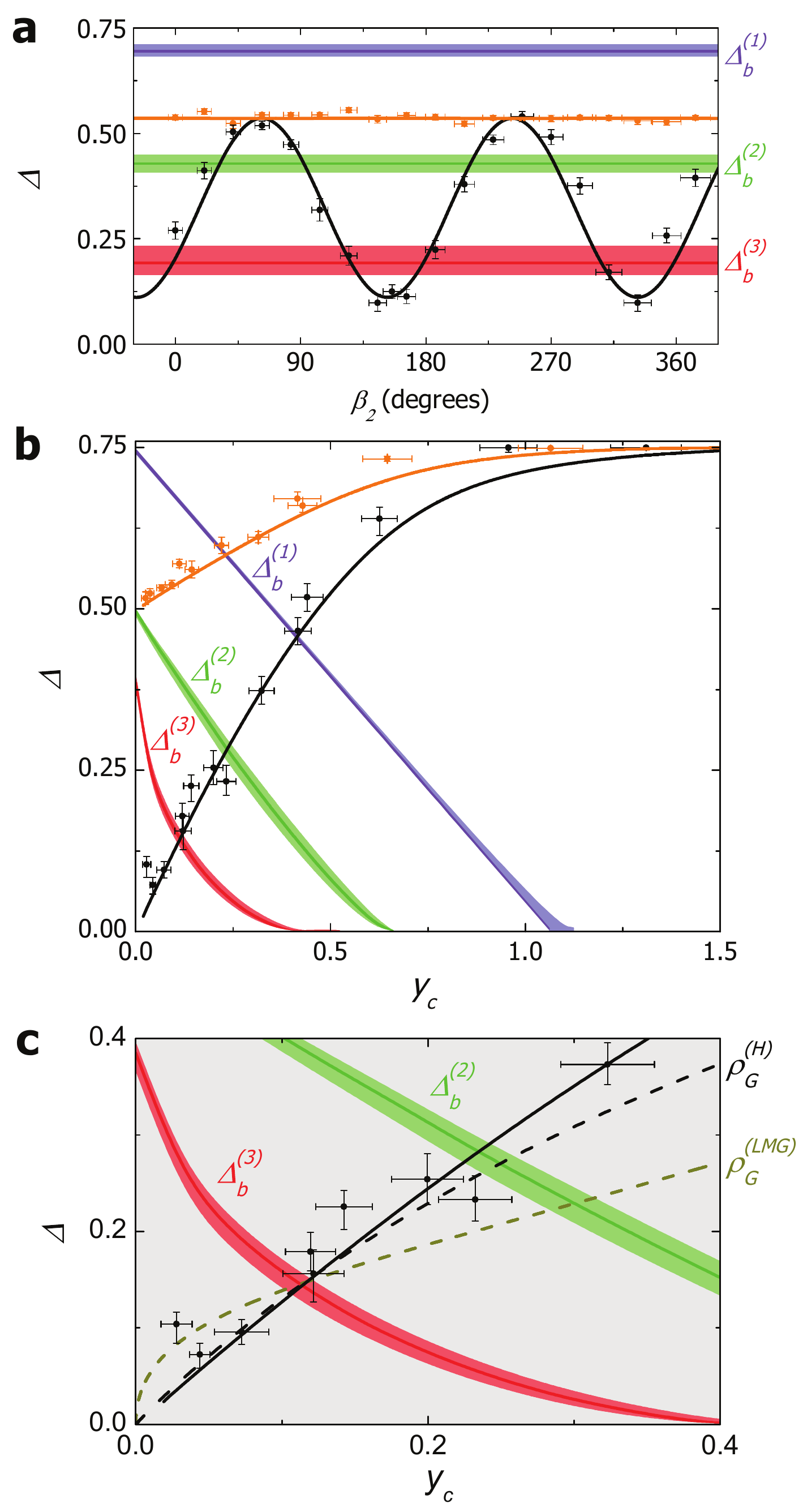}
\caption{\fontfamily{ptm}\selectfont\small\textbf{Quadripartite entanglement among four atomic ensembles.} \textbf{a,} Phase coherence between the bipartite entangled pairs of the quadripartite state. We observe an interference fringe between the bipartite modes $\{a_{2},b_{2}\}$ and  $\{c_{2},d_{2}\}$ of the full quadripartite state (black points) as a function of bipartite phase $\beta_2$. As a control experiment, we also generate a `crossed' quantum state $\hat{\rho}^{(A)}_X$ whose bipartite components for the split between $\{a,b\}$ and $\{c,d\}$ are in principle distinguishable (orange points and line). \textbf{b,} Exploring the entanglement space $\{\Delta,y_c\}$ for quadripartite states. By controlling the spin-wave statistics for the state of the four ensembles, we observe transitions from quadripartite, to tripartite, to bipartite entangled states, and to fully separable states (black points). Results for the `crossed' quantum state $\hat{\rho}^{(A)}_X$ are shown by the orange points and line. \textbf{c,} Expanded view of entanglement parameters near the origin for $\{\Delta,y_c\}$. We also display results for entanglement thermalization $\{\Delta^{(T)},y_c^{(T)}\}$ of the spin systems $\hat{\rho}^{(\text{H})}_G$ ($\hat{\rho}^{(\text{LMG})}_G$) by the black (olive) dashed line. The red, green and purple bands represent the minimum uncertainties for three-mode ($\Delta^{(3)}_b$) and two-mode entanglement ($\Delta^{(2)}_b$), and for fully separable states ($\Delta^{(1)}_b$). In all cases, error bars for the data points reflect the statistical and systematic uncertainties.
The thicknesses of the $\{\text{red, green, purple}\}$ bands from the central lines correspond to the $\pm 1$-s.d. uncertainties of the bounds $\{\Delta^{(3)}_{b},\Delta^{(2)}_{b},\Delta^{(1)}_{b}\}$, due to losses and imbalances of the verification interferometer.}
\label{fig2}
\end{figure}

Figure \ref{fig2} presents our results for quadripartite entanglement achieved for storage time $\tau_0 =0.2$ $\mu \text{s}$. By appropriate settings of the interferometric phases for the writing and heralding processes in step (1), we prepare a correlated atomic state where the coherent component $\hat{\rho}^{(A)}_c$ of $\hat{\rho}^{(A)}_W$ is established with well-defined phase relations among the four ensembles $\{a,b,c,d\}$ as for the ideal $W$-state in Eq. \ref{atomWstate}. After a delay $\tau_0$ for step (2) quantum storage, the atomic state $\hat{\rho}^{(A)}_W(\tau_0)$ is mapped to the photonic state $\hat{\rho}^{(\gamma)}_W(\tau_0)$ in step (3) by way of read pulses. For step (4), the phases $\{\beta_1,\beta_2,\beta_3\}_v$ of the verification interferometer are optimized to match the orientations of the projectors $\{\hat{\Pi}^{(c)}_i\}$ to the stable photonic phases $\{\phi_1^{\prime},\phi_2^{\prime},\phi_3^{\prime}\}$ of $\hat{\rho}^{(\gamma)}_W$. The photon probabilities for $q_{ijkl}$ and $p_{ijkl}$ are then recorded as in the upper and lower panels of Fig. \ref{fig1}c.

We first investigate off-diagonal coherence for the purportedly entangled atomic and photonic states, $\hat{\rho}^{(A)}_{W}(\tau)$ and $\hat{\rho}^{(\gamma)}_{W}(\tau)$, in Fig. \ref{fig2}a. As the bipartite phase $\beta_2$ is varied, we observe interference fringes in the outputs $\{p_{1000},p_{0100},p_{0010},p_{0001}\}$ of the verification interferometer, and hence a variation in $\Delta$, that results from the coherence between the bipartite entangled components of $\hat{\rho}^{(\gamma)}_{W}(\tau)$ for the modes $\{a_2,b_2\}$ and $\{c_2,d_2\}$. Furthermore, for optimal phase settings of $\beta_2$, the observed values of $\Delta$ fall below the bounds $\{\Delta^{(3)}_b,\Delta^{(2)}_b,\Delta^{(1)}_b\}$ (red, green, purple bands) for the measured value of $y_c$ and thereby certify that a fully quadripartite entangled state $\hat{\rho}^{(\gamma)}_W$ (and thereby $\hat{\rho}^{(A)}_W$) has been generated. In particular, this state verifiably possesses stronger quantum correlations than any admixtures of tripartite entangled ($\Delta < \Delta^{(3)}_b$), bipartite entangled ($\Delta < \Delta^{(2)}_b$), and fully separable states ($\Delta < \Delta^{(1)}_b$). Quantitatively, we find the uncertainty violation $\Delta=0.10\pm 0.01 <\Delta^{(3)}_b=0.19 ^{+0.05}_{-0.01} $ at $y_c=0.06\pm 0.02$ and $\beta_2=175^{\circ}$. This is in reasonable correspondence to the result $\Delta_{\text{th}}=0.08$ from our theoretical model.

The quadripartite entanglement observed in Fig. \ref{fig2}a arises from the intrinsic indistinguishability of probability amplitudes for creation of one collective excitation $|\overline{s}\rangle_{\epsilon}$ among the four ensembles, as implemented by the heralding measurement $\hat{\Pi}_h$ for fields $\gamma_1=\{a_{1},b_{1},c_{1},d_{1}\}$. As a control experiment, we reconfigure the heralding interferometer such that the fields $\{a_{1},b_{1}\}$ and $\{c_{1},d_{1}\}$ are combined with orthogonal polarizations ($\text{BS}_X$) immediately before the heralding detector $D_h$ in Fig. \ref{fig1}b. In this case, path-information could in principle be revealed up to the bipartite split of the ensembles $\{a,b\}$ and $\{c,d\}$ by discriminating the polarization state of the heralding photon. Therefore, the heralding measurement $\hat{\Pi}_X$  prepares a `crossed' atomic state $\hat{\rho}^{(A)}_{X}$ with no coherence shared between the ensembles $\{a,b\}$ and $\{c,d\}$. We observe an absence of interference as $\beta_2$ is varied in Fig. \ref{fig2}a (orange points). However, this modified heralding process $\hat{\Pi}_X$ preserves bipartite entanglement within the ensemble pairs $\{a,b\}$ and $\{c,d\}$, which explains our observation of the uncertainty $\Delta= 0.533\pm0.005$ (close to the predicted result $\Delta^{(X)}_{\text{th}}=0.536$ (orange line)) reduced below the 1-mode bound $\Delta^{(1)}_b=0.70^{+0.01}_{-0.02}$ for $y_c=0.07\pm 0.01$, thereby supporting the presence of bipartite entanglement for the ensemble pairs  $\{a,b\}$ and $\{c,d\}$. 

Next, we characterize $\hat{\rho}^{(\gamma)}_W$ (and thereby $\hat{\rho}^{(A)}_W$) over the full parameter space of $\{\Delta,y_c\}$. In a regime of weak excitation for each of the ensemble-field pairs $\{\epsilon,\gamma_1\}$, with $\epsilon=\{a,b,c,d\}$ and $\gamma_{1}=\{a_{1},b_{1},c_{1},d_{1}\}$, the heralded state $\hat{\rho}^{(A)}_{W}$ includes small, higher-order components and is approximately given by
\begin{equation}
\label{atomWrho}
\hat{\rho}^{(A)}_{W}(\tau=0)\simeq  (1-3\xi)| W \rangle_{_{A}}\langle W |  + 3\xi \hat{\rho}^{(A)}_{\geq 2}+O(\xi^2),
\end{equation}
where $\hat{\rho}^{(A)}_{\geq2}$ includes spin-waves with two or more quanta in the set of four ensembles as well as uncorrelated excitations due to atomic noise.  For excitation probability $\xi\rightarrow 0$, a heralding event at $D_h$ leads to a state with high fidelity to $| W \rangle_{_{A}}$ in Eq. \ref{atomWstate} stored in the four ensembles. However, for increasing $\xi$, higher-order terms with multiple excitation number become important, leading to modifications of the spin-wave statistics for the heralded state $\hat{\rho}^{(A)}_{W}$ and thereby to the entanglement parameters $\{\Delta,y_c\}$. For $\xi\ll 1$, we find theoretically that $(y_{c})_{\text{th}}\simeq 8\xi$ and $\Delta_{\text{th}}\simeq 9\xi$ (Eq. \ref{atomWrho}), excluding any external noise sources and assuming perfect spatio-temporal overlaps for the quantum fields $\gamma_{2}=\{a_{2},b_{2},c_{2},d_{2}\}$ in the verification interferometer. Hence, by varying the excitation probability $\xi$ via the overall intensity for the write beam, we can simultaneously adjust the statistical contamination $y_c$ and sum uncertainty $\Delta$ of the entangled states $\{\hat{\rho}^{(A)}_W,\hat{\rho}^{(\gamma)}_W\}$ across their entanglement spaces.

This procedure is employed in Fig. \ref{fig2}b to parametrically increase $\{\Delta,y_c\}$ in tandem by increasing $\xi$. As the statistical contamination $y_c$  is raised from $y_c\simeq 0$ in the quantum domain of one-excitation to the classical regime of multiple excitations with $y_c\simeq 1$, we observe the transitions of the directly measured photonic $W$-states $\hat{\rho}^{(\gamma)}_W$ (black points) from fully quadripartite entangled to tripartite entangled (from $\Delta<\Delta_b^{(3)}$ to $\Delta>\Delta_b^{(3)}$), to bipartite entangled (from $\Delta<\Delta_b^{(2)}$ to $\Delta>\Delta_b^{(2)}$), and finally to fully separable states (from $\Delta<\Delta_b^{(1)}$ to $\Delta>\Delta_b^{(1)}$). 
In comparison to our former work on photonic $W$-states via coherent splitting of a photon \cite{papp09}, the heralded atomic and photonic $W$-states $\{\hat{\rho}^{(A)}_{W},\hat{\rho}^{(\gamma)}_{W}\}$ offer qualitatively richer statistical passages through the entanglement spaces $\{\Delta,y_c\}$. Here, the quantum coherence of $\{\hat{\rho}^{(A)}_W,\hat{\rho}^{(\gamma)}_W\}$ evidenced by $\Delta$ is inherently linked to the statistical character of $\{\hat{\rho}^{(A)}_{W},\hat{\rho}^{(\gamma)}_{W}\}$ expressed by $y_c$ due to increasing decorrelations between the heralding fields $\gamma_1$ and the excitation statistics of the ensembles with increasing $\xi$. 

For $\xi\ll 1$, the coherent contribution $\hat{\rho}^{(A)}_c$ of the delocalized single quantum strongly dominates over any other processes of the full quadripartite state $\hat{\rho}^{(A)}_W$ including vacuum components and multiple spin-wave excitations arising from the uncorrelated excitations of $\hat{\rho}^{(A)}_{\geq 2}$ (Eq. \ref{atomWrho}). With a heralding probability $p_h\simeq 3\times 10^{-4}$ (corresponding to $\xi\simeq 5\times 10^{-3}$), we achieve the smallest entanglement parameters of $\Delta^{\text{min}}=0.07^{+0.01}_{-0.02}$ and $y_c^{\text{min}}=0.038\pm 0.006$ for the generated quadripartite entangled states. These entanglement parameters are suppressed below the closest 3-mode boundary ${\Delta}_{b}^{(3)}$ by more than ten standard deviations (s.d.). Furthermore, because the mapping of quantum states from matter to light cannot increase entanglement\cite{chou05}, our measurements of $\hat{\rho}_W^{(\gamma)}$ unambiguously provides a lower bound of the quadripartite entanglement stored in $\hat{\rho}_W^{(A)}$. Thereby, the observed strong violation of the uncertainty relations for $\{\Delta^{\text{min}},y_c^{\text{min}}\}$ categorically certifies for the creation of measurement-induced entanglement of spin-wave excitations among four quantum memories, as well as for the faithful and coherent transfer of the stored quadripartite entangled states to an entangled state of four propagating electromagnetic fields. We emphasize that our analysis makes use of the full physical state $\{\hat{\rho}^{(\gamma)}_W,\hat{\rho}^{(A)}_W\}$ including the vacuum component $\hat{\rho}_{0}$ and higher order terms $\hat{\rho}_{\geq 2}$, and does not rely upon a spurious post-diction based upon a preferred set of detection events. 

As shown by the black and orange curves in Fig. \ref{fig2}, our observations correspond well to a theoretical model of the heralding $\hat{\rho}^{(A)}_{W}= \text{Tr}_{h}(\hat{\Pi}_{h}\hat{U}^{\dagger}_{\text{write}}\hat{\rho}^{(A)}_{g}\hat{U}_{\text{write}})$, readout $\hat{\rho}^{(\gamma)}_{W}=\text{Tr}_A( \hat{U}^{\dagger}_{\text{read}}\hat{\rho}^{(A)}_{W}\hat{U}_{\text{read}})$, and verification steps of our experiment, with $\hat{\rho}^{(A)}_{g}=|\overline{g}\rangle\langle\overline{g}|$. In addition to Eq. \ref{atomWrho}, the model includes the effects of atomic fluorescence, background noises, and finite efficiencies, as well as various imperfections in the measurements.

\begin{figure*}[tH!]
\includegraphics[width= 1.8\columnwidth]{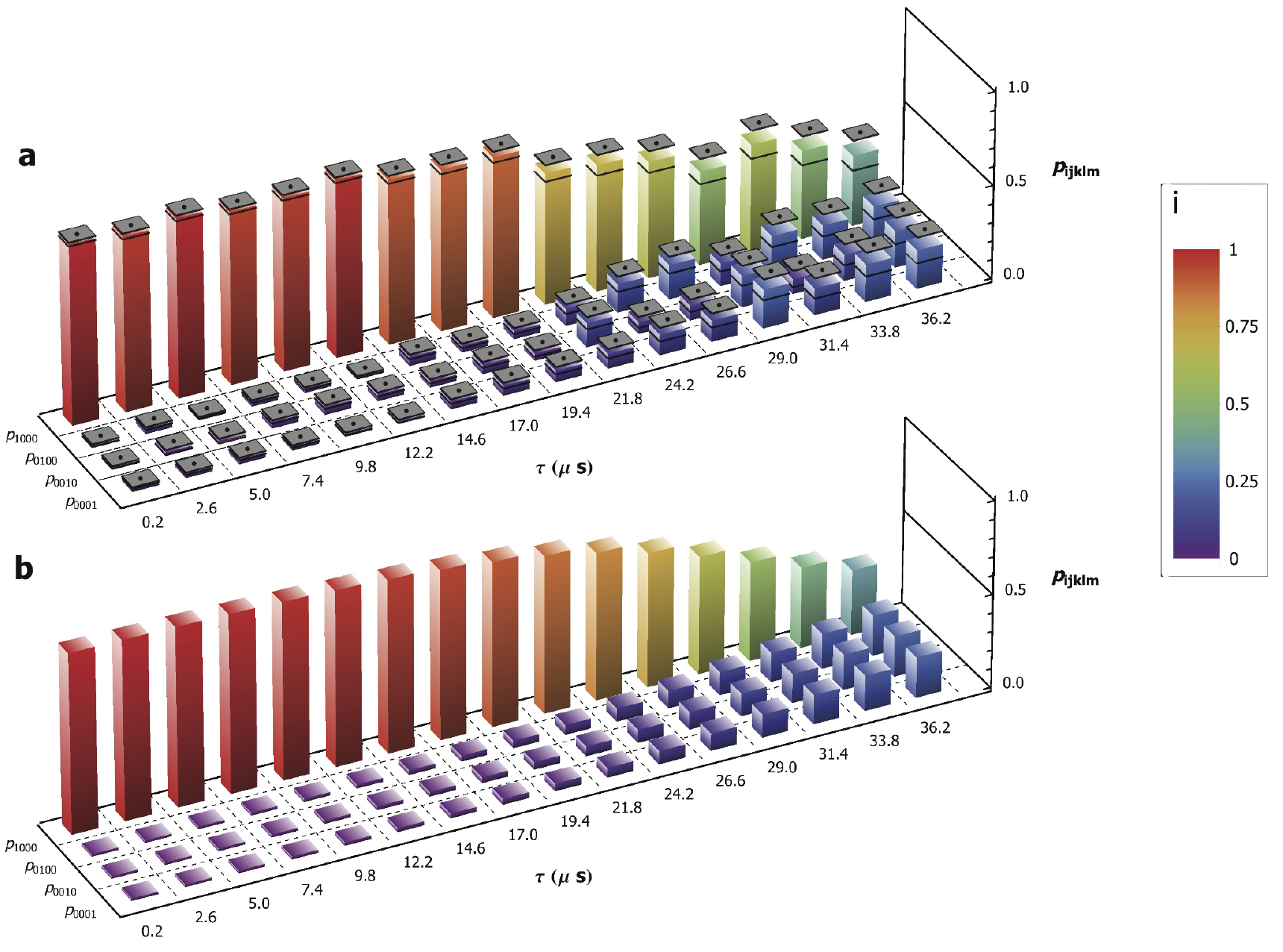}
\caption{\fontfamily{ptm}\selectfont \small\textbf{Temporal decay of coherences stored in four atomic ensembles.} \textbf{a,} Evolution of the photon probabilities $\{p_{1000},p_{0100},p_{0010},p_{0001}\}$ for occupying the output modes of the verification interferometer versus storage time $\tau$. For readability, the heights of the bars are shown in accord to the color convention of inset \textbf{i}. Error bars, shown as grey squares, reflect the statistical uncertainties for each point. \textbf{b,} Photon probabilities $\{p_{1000},p_{0100},p_{0010},p_{0001}\}$ from our theoretical model, which assumes a memory time determined from the temperature of the cold atomic samples and the net momentum transfer to the atomic spin-waves (Appendix).}
\label{fig3}
\end{figure*}

In terms of state fidelity, our approach for multipartite entanglement generation compares favorably to matter systems utilizing local interactions (e.g., trapped ions \cite{leibfried05a,haffner05a,barreiro10}). Instead of direct interactions, the atomic $W$-state in our experiment is generated in a heralded fashion by a quantum measurement mediated by the initial atom-photon $\{\epsilon,\gamma_1\}$ correlations\cite{duan01,chou05}. Despite the intrinsically low preparation probability $p_h$ and losses in the heralding channels, the resulting quadripartite entangled state $\hat{\rho}_W^{(A)}$ stored in the four ensembles has high fidelity with the ideal $W$-state in Eq. \ref{atomWstate}, namely  $F^{(A)}=\langle W_A| \hat{\rho}^{(A)}_W | W_A \rangle$. Indeed, we derive a fidelity $F^{(A)}_{\text{th}}=1-3\xi$ from Eq. \ref{atomWrho}, and thereby deduce a theoretical fidelity $F^{(A)}_{\text{th}}=0.98$. Experimentally, because the measurements $\{\hat{\Pi}_i^{(c)}\}$ of the sum uncertainty project the input state $\hat{\rho}_{r}$ into the four orthonormal $W$-states $|W_i\rangle_v$, we can attribute a lower bound for the entanglement fidelity $\tilde{F}^{(A)}=\langle W_1|\hat{\rho}^{(A)}_W | W_1\rangle=\tilde{p}_{1} F_1^{(A)}$ for the heralded atomic state $\hat{\rho}^{(A)}_W$, with $|W_1\rangle\in\{|W_i\rangle_v\}$. Specifically, the single-excitation probability, $\tilde{p}_1$, and the conditional fidelity, $F_1^{(A)}=\langle W_1| \hat{\rho}^{(A)}_1 | W_1   \rangle$, for the atomic ensembles are determined from the respective inferences of the spin-wave statistics (via the retrieval efficiency $\eta_{\text{read}}$ and $y_c$), and of the coherences (via $\Delta$). 
For the parameters  $\{\Delta^{\text{min}},y_c^{\text{min}}\}$ and $\eta_{\text{read}}=38\pm 4\%$, we deduce an entanglement fidelity $\tilde{F}^{(A)}=\tilde{p}_{1} (\sqrt{\frac{1}{2}(\frac{1}{2}-\Delta)}+\frac{1}{2}) =0.9\pm 0.1 \simeq F^{(A)}_{\text{th}}$ for the stored atomic state (Appendix).

Apart from the creation of a novel multipartite entangled state of spin-waves with tunable quantum statistics, an important benchmark of our quantum interface is the transfer efficiency $\lambda$ of the quadripartite $W$-states from matter to light\cite{choi08}. Since no known measure applies to our case, we tentatively define the entanglement transfer $\lambda =\tilde{F}^{(\gamma)}/\tilde{F}^{(A)}$  by the ratio of the respective physical fidelities $\tilde{F}^{(A)}$ and $\tilde{F}^{(\gamma)}$ for the atomic and photonic states. In particular for $\xi\ll 1$, the photonic fidelity can be approximated by $\tilde{F}^{(\gamma)}_{\text{th}}\simeq\eta_{\text{read}}\tilde{F}^{(A)}_{\text{th}}$, which thereby gives $\lambda_{\text{th}}\simeq\eta_{\text{read}}$ dictated by the retrieval efficiency $\eta_{\text{read}}$. For the minimal entanglement parameters $\{\Delta^{\text{min}},y_c^{\text{min}}\}$ in Fig. \ref{fig2}b, we find $\tilde{F}^{(\gamma)}=0.35\pm 0.02$ and obtain an entanglement transfer $\lambda=0.39\pm0.04$, similar to the observed $\eta_{\text{read}}$. While fidelity is an often used measure, we emphasize that $\tilde{F}^{(\gamma)}$ cannot be used to set a threshold for entanglement for $\hat{\rho}_W^{(\gamma)}$, since $\hat{\rho}_W^{(\gamma)}$ can exhibit multipartite entanglement for any $\tilde{F}^{(\gamma)}>0$. 

\begin{figure*}[t]
\includegraphics[width= 2\columnwidth]{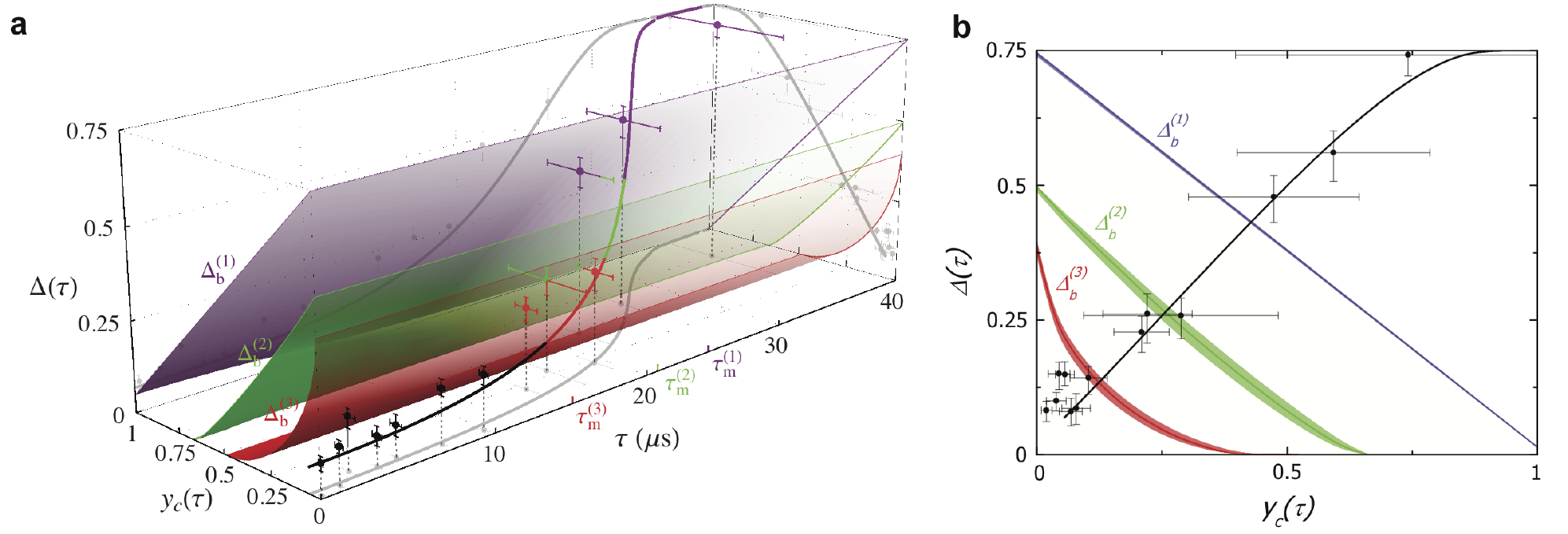}
\caption{\fontfamily{ptm}\selectfont \small\textbf{Dissipative dynamics of atomic entanglement.} \textbf{a,} Time-evolution of the entanglement parameters $\{\Delta(\tau),y_c(\tau)\}$ versus $\tau$ for the multipartite quantum state. We observe the crossing of the boundaries defining the minimum uncertainties for 3-mode (red surface, $\Delta_b^{(3)}$), 2-mode (green surface, $\Delta_b^{(2)}$) entangled states, and 1-mode (purple surface, $\Delta_b^{(1)}$) separable states. For clarity, the data points and theoretical curve are colored to indicate the various entanglement orders for the quadripartite (black), tripartite (red), bipartite entangled (green) states, and fully separable states (purple). The projections of the data points into the planes $(y_c,\tau)$ and $(\Delta,\tau)$ are shown as gray points to display the individual passages of the quantum statistics and coherences, respectively. \textbf{b,} Projection of entanglement dynamics onto the $(\Delta,y_c)$ plane. The various  entanglement transitions from \textbf{a} are shown for our measurement and theory. Error bars for the data represent $1/e$-errors from a conservative optimization analysis for evaluating $\{\Delta,y_c\}$ that reflects the parameters' statistical and systematic uncertainties. The curves are from a theoretical model that includes motional dephasing.}
\label{fig4}
\end{figure*}

To investigate the dynamical behavior of the observed quadripartite entangled states, we study the temporal evolution of multipartite entanglement stored in the atomic ensembles as a function of time $\tau$ following the heralding event at $\tau=0$. Decoherence for the atomic $W$-state is governed by independent dephasing processes in each ensemble, in which the imprinted atomic phases in the collective state $|\overline{s}_{\epsilon}\rangle$ evolve independently due to atomic thermal motion, which transforms the initial collective atomic state into a subradiant state\cite{simon07b} that becomes increasingly uncorrelated with the heralding fields $\gamma_1=\{a_{1},b_{1},c_{1},d_{1}\}$, for each ensemble $\epsilon =\{a,b,c,d\}$. The subradiant state contributes to a reduction in the coherent component $\hat{\rho}_c^{(A)}$ of $\hat{\rho}_W^{(A)}$ as well as to a build-up of uncorrelated atomic noise $\hat{\rho}^{(A)}_{\geq 2}$ relative to $\hat{\rho}_c^{(A)}$. The net effect is an increase of both entanglement parameters $\{\Delta,y_c\}$ with a typical time-scale of $\tau_m=1/(|\delta{\vec{k}}|  v_{d})\simeq 17$ $\mu s$, where $\delta{\vec{k}}$ is the momentum transfer to the spin-waves and $v_d$ is the average velocity of the atoms\cite{simon07b}. Eventually, the growth in $\{\Delta(\tau),y_c(\tau)\}$ leads to time-dependent losses of entanglement, marked by successive crossings of the boundaries set by $\{\Delta_b^{(3)},\Delta_b^{(2)},\Delta_b^{(1)}\}$. 

Fig. \ref{fig3} illustrates the temporal reduction in the overall coherence $\overline{d}$ of the full quadripartite state. Operationally, the loss of coherence is observed in terms of decrease in imbalances among $\{p_{1000},p_{0100},p_{0010},p_{0001}\}$ as a function of storage time $\tau$, and hence to an increase in $\Delta$. The behavior of the experimentally observed photon probabilities in Fig. \ref{fig3}a results from the progressive decay of the initial coherence for $\hat{\rho}_{W}^{(A)}(\tau_i)$ at $\tau_{i}=0.2$ $\mu s$ for which $V_{\text{eff}}(\tau_i)=4\overline{d}=0.95\pm 0.02$, evolving then to $V_{\text{eff}}=0.10^{+0.25}_{-0.10}$ for the final state $\hat{\rho}_{W}^{(A)}(\tau_f)$ measured at $\tau_f=36.2$ $\mu$s. The observed evolution is in good agreement with our theoretical model of the photon probabilities shown in Fig. \ref{fig3}b. The spin-wave statistics are similarly modified by phase decoherence leading to an increase of $y_c$, from $y_c(\tau_i)=0.03\pm 0.01$ to $y_c(\tau_f)=0.74\pm 0.34$.

Finally, in Fig. \ref{fig4} we examine the dissipative dynamics of entanglement for the quantum memories of four ensembles via the evolution of both $\{\Delta,y_c\}$. We observe the passage of the initial quadripartite entangled state $\hat{\rho}^{(A)}_W (\tau_i)$ through various domains, evolving from $M$-partite entanglement to $(M-1)$-partite entanglement at memory times $\tau=\tau_{m}^{(M-1)}$. The crossings of the bounds $\{\Delta_b^{(3)},\Delta_b^{(2)},\Delta_b^{(1)}\}$ occur at $\tau_{m}^{(3)}=15$ $\mu s$, $\tau_{m}^{(2)}=21$ $\mu s$, and $\tau_{m}^{(1)}= 24$ $ \mu s$, respectively. In addition, we find that the measured entanglement parameters evolve in accord to the simulated dynamics derived for $\hat{\rho}^{(A)}_W(\tau)$ from our theoretical model (solid line). In Fig. \ref{fig4}b we show the projection of Fig. \ref{fig4}a into the plane of $\{\Delta,y_c\}$, which directly reveals the losses of entanglement, albeit without any temporal information.

An interesting extension of our work is to relate the characterization of multipartite entanglement by way of $\{\Delta,y_c\}$ as in Fig. \ref{fig2} to the thermal relaxation of entanglement for quantum many-body systems\cite{amico08,guhne09}. We consider a ferromagnetic ($J>0$) Heisenberg-like model $\hat{H}_{\text{H}}^{\prime}=\hat{H}_{\text{H}}+\hat{H}_p$ and a Lipkin-Meshkov-Glick model $\hat{H}_{\text{LMG}}=-\frac{J}{4}\sum_{\langle i,j \rangle}(\hat{S}^{(i)}_x\cdot\hat{S}^{(j)}_x+\hat{S}^{(i)}_y\cdot\hat{S}^{(j)}_y)+h_z\sum_i \hat{S}_z^{(i)}$, for four spins $\{\hat{S}^{(i)}_x,\hat{S}^{(i)}_y,\hat{S}^{(i)}_z\}$ on a tetrahedron. Here, $\hat{H}_{\text{H}}^{\prime}$ includes the standard Heisenberg Hamiltonian $\hat{H}_{\text{H}}=-\frac{J}{4}\sum_{\langle i,j\rangle}{{\vec{S}}^{(i)} \cdot{\vec{S}}^{(j)}}+h_z \sum_{i}{S}_z^{(i)}$, in addition to a spin-projection term $\hat{H}_p=2h_z|S,-S\rangle\langle S,-S|$, which shifts the energy of the collective spin state $|S,-S\rangle$. Formally, the entangled state $\hat{\rho}_W^{(A)}$ in Eq. \ref{atomWrho} closely resembles the thermal equilibrium state $\hat{\rho}_{_{G}}^{(\text{H})}$ ($\hat{\rho}_{_{G}}^{(\text{LMG})}$) of $\hat{H}_{\text{H}}^{\prime}$ ($\hat{H}_{\text{LMG}}$). The eigenstates of $\hat{H}_{\text{H}}^{\prime}$ and $\hat{H}_{\text{LMG}}$ are the total angular momentum states defined by collective spin algebra, whose ground state at zero temperature ($T\rightarrow 0$) is an entangled $W$-state similar to Eq. \ref{atomWstate} for magnetic field $h_z=J/2$. By redefining $\{|\downarrow\rangle,|\uparrow\rangle\}\mapsto\{|0\rangle,|1\rangle\}$, we calculate the Gibbs state $\hat{\rho}_{_{G}}^{(\text{H})}=(1/Z)e^{-\hat{H}_{\text{H}}^{\prime}/k_B T}$ ($\hat{\rho}_{_{G}}^{(\text{LMG})}=(1/Z)e^{-\hat{H}_{\text{LMG}}/k_B T}$) with partition function $Z$ as a function of $\{\Delta^{(T)},y_c^{(T)}\}$ derived for this system $\hat{H}_{\text{H}}^{\prime}$ ($\hat{H}_{\text{LMG}}$). The results are shown in Fig. \ref{fig2}c by the black (olive) dashed lines for $\hat{\rho}_{_{G}}^{(\text{H})}$ ($\hat{\rho}_{_{G}}^{(\text{LMG})}$). The statistical character of $\hat{\rho}_W^{(A)}$ (black points) for our system of four ensembles follows closely the thermalization process relevant for the low-lying states of $\hat{H}_{\text{H}}^{\prime}$ (black dashed line) for $y_c\lesssim 0.2$, whereby the higher-order spin-waves $\hat{\rho}^{(A)}_{\geq 2}$ are populated by thermal excitations. While intriguing, such an analogy is incomplete for our system, because $\hat{\rho}^{(A)}_W$ is comprised of four $N_{A,\epsilon}$ dimensional quasi-bosonic excitations instead of the four single-spins $\{|\downarrow\rangle,|\uparrow\rangle\}$ of $\hat{\rho}_{_{G}}^{(\text{H})}$, $\hat{\rho}_{_{G}}^{(\text{LMG})}$. Nevertheless, our methodology for multipartite entanglement characterization is clearly applicable to quantum spin systems and may help to advance our understanding of the link between off-diagonal long-range order and multipartite entangled spin-waves in quantum magnets \cite{amico08,guhne09}.

In conclusion, our measurements explicitly demonstrate a coherent matter-light quantum interface for multipartite entanglement by way of the operational metric of quantum uncertainty relations \cite{hofmann03,papp09,lougovski09}. High-fidelity entangled spin-waves are generated in four spatially separated atomic ensembles and subsequently coherently transferred to quadripartite entangled beams of light. The quantum memories provided by the four atomic ensembles are individually addressable and can be readily read-out at different times for conditional control of entanglement distribution and connection\cite{duan01}. By way of recent advances from other groups (Appendix), the short memory times for our work could be significantly improved beyond one second. In addition, measurement-induced generation of entanglement could be augmented by the reversible mapping of entangled photonic states to matter \cite{choi08}.

Other possibilities include the creation of yet larger multipartite entangled states with efficient scaling \cite{duan01}. For example, quadripartite entangled states of ensemble sets $\{a,b,c,d\}$ and $\{a^{\prime},b^{\prime},c^{\prime},d^{\prime}\}$ could be extended by swapping between ensembles $\{a,a^{\prime}\}$ to prepare a hexapartite (6-partite) entangled state for ensembles $\{b,b^{\prime},c,c^{\prime},d,d^{\prime}\}$ (see Appendix). Generalization of such entanglement swapping processes may help to prepare a single macroscopic entangled state which fully occupies an entire network, for exploratory goals of observing the extreme non-locality of $W$-states \cite{heaney10} and entanglement percolation\cite{acin07}. In addition, the expansion of multipartite quantum networks with increasing $N$ offers a promising bottom-up approach for the study of phase transitions in strongly-correlated systems \cite{amico08,guhne09}. Finally, the entangled spin-waves comprising the network can be used to implement a quantum-enhanced parameter estimation scheme for detecting an unknown phase shift of $\pi$ for one component of $\hat{\rho}_W^{(A)}$ with efficiency beyond any separable state, which could find application to quantum metrology.

\newpage\newpage
\noindent\textbf{Appendix}

\noindent\textbf{Experimental details.} 
The experiment consists of a $22$ ms preparation stage and a $3$ ms period for operating the quantum interface in Fig. \ref{fig1} with a repetition rate of $R_{l}=40$ Hz. For the preparation, we load and laser-cool Caesium atoms (peak optical depth $\approx 30$) in a magneto-optical trap for $18$ ms, after which the atoms are further cooled by optical molasses ($T_{d}\simeq 150$ $\mu K$) and optically pumped to $|g\rangle$ for $4$ ms. During this time, a phase-reference laser ($F=3 \leftrightarrow F^{\prime}=4^{\prime}$ transition) also propagates through the atomic ensembles for the active stabilization of the verification interferometer in Fig. \ref{fig1}c via \textit{ex-situ} phase-modulation spectroscopy\cite{papp09}, which does not affect the operation of the quantum interface. Concurrently, dense Caesium atoms in paraffin coated vapor cells located at the heralding and verification ports are prepared to the respective ground states $|g\rangle$ ($|s\rangle$) for filtering the classical lasers scattered into the quantum fields $\gamma_1$ ($\gamma_2$). 

\noindent\textbf{Quantum interface.} 
For the quantum interface to function during the $3$ ms window, in step (1) $20$-ns wide writing (red-detuned $\delta=10 \text{ MHz}$ from $|g\rangle - |e\rangle$ transition)  and reading pulses (resonant with $|s\rangle - |e\rangle$) are applied sequentially to the ensembles $\epsilon$, synchronized to a clock running at $R_{c}\simeq 2$ MHz. This process creates pairwise correlated excitations\cite{duan01} between the collective atomic modes $|\overline{s}\rangle_{\epsilon}$ of the ensembles $\epsilon$ and the optical fields $\gamma_1$ ($\delta=10 \text{ MHz}$ below $|s\rangle - |e\rangle$). Photodetection of a single photon for the combined fields $\gamma_1$ at the output of the heralding interferometer effectively erases the which-path information for $\gamma_1$, and imprints the entangled spin-wave $\hat{\rho}^{(A)}_{W}$ (Eq. \ref{atomWrho}) onto the ensembles $\{a,b,c,d\}$ via $\text{Tr}_{h}(\hat{\Pi}_{h}\hat{U}^{\dagger}_{\text{write}}\hat{\rho}^{(A)}_{g}\hat{U}_{\text{write}})$. The heralding event at $D_{h}$ triggers control logic in Fig. \ref{fig1}a which deactivates intensity modulators of the write (IM$_{\text{write}}$) and read lasers (IM$_{\text{read}}$) for the quantum storage of $\hat{\rho}^{(A)}_{W}$ in step (2). After a user-controlled delay $\tau$, step (3) is initiated with strong read pulses (Rabi frequency $24$ MHz) that address the ensembles in Fig. \ref{fig1}c and coherently transfer the entangled atomic components  $\{a,b,c,d\}$ of $\hat{\rho}^{(A)}_{W}(\tau)$ one-by-one to propagating beams $\gamma_2 =\{a_{2},b_{2},c_{2},d_{2}\}$, comprising the entangled photonic state $\hat{\rho}^{(\gamma)}_{W}(\tau)$, via the operation $\hat{\rho}^{(\gamma)}_{W}=\text{Tr}_A( \hat{U}^{\dagger}_{\text{read}}\hat{\rho}^{(A)}_{W}\hat{U}_{\text{read}})$. Here, $\text{Tr}_A$ traces over the atomic systems which are later shelved into the ground states $|\overline{g}\rangle$. The retrieval efficiency $\eta_{\text{read}}$ is collectively enhanced for large $N_{A}$ (ref. \cite{duan01}), leading to $\eta_{\text{read}}=0.38\pm0.06$ in our experiment.

\noindent\textbf{Spin-wave quantum memories.}
The quantum information of the entangled state for Eq. \ref{atomWstate} is encoded in the quantum numbers of spin-waves (collective excitations) for the pseudo-spin of the hyperfine ground electronic levels $6S_{1/2},F=3,F=4$ in atomic Caesium. The fluorescence images shown in the inset of Fig. \ref{fig1}a arise from excitation by the writing and reading beams with $1$ $\text{mm}$ separations and $200$ $\mu m$ waists. The geometry of actual collective excitations for the four ensembles $\{a,b,c,d\}$ is defined by the spread functions of the imaging systems for the fields $\gamma_1,\gamma_2$, which have waists of $60$ $\mu m$ for each ensemble consisting of a cold cloud of $N_{A,\epsilon}\approx 10^6$ Caesium atoms. We use an off-axial configuration\cite{balic05} for addressing each ensemble $\epsilon$ with an angle $\theta=2.5 ^{\circ}$ between the classical and nonclassical beams, that creates spin-waves $|\overline{s}\rangle_{\epsilon}$ associated with wave-numbers $\delta \vec{k}=\vec{k}_{\text{write}}-\vec{k}_{\gamma_1}$ for each $\epsilon$. These spin-waves are analogous to other types of collective excitations in many-body systems, such as magnons and plasmons, and the spin-waves can be converted to dark-state polaritons for the coherent transfer $\hat{U}_{\text{read}}$ of entanglement. For the phase-matching configuration and temperature of our ensembles, the memory times $\{\tau^{(3)}_{m},\tau^{(2)}_{m},\tau^{(1)}_{m}\}$ in Fig. \ref{fig4} are dominantly determined by the motional dephasing of the spin-waves $|\overline{s}\rangle_{\epsilon}$ \cite{simon07b}. With thermal velocity of $v_d\simeq 14$ $\text{cm}/\text{s}$, we theoretically determine a memory time $\tau_m=\frac{0.85\mu m}{4 \pi {sin}(\theta/2) v_d}=17$ $\mu s$. On the other hand, the ground-state dephasing due to inhomogeneous broadening is expected to be $>50$ $\mu s$ in our experiment, inferred from two-photon Raman spectroscopy.

\noindent\textbf{Derivation of entanglement fidelity.}
We derive here the expression for the lower bound entanglement fidelity $\tilde{F}^{(A)}=\tilde{p}_{1} F_1$. We start by noting that the projective measurement $\hat{\Pi}^{(c)}_{i}$ for $\Delta$ gives the conditional fidelity $F_1$ of the input $\hat{\rho}_r$ onto one of four orthonormal $W$-states, $|W_i\rangle_v=|W_1\rangle_v$, for example,  $|1000\rangle+e^{i\beta_1}|0100\rangle+e^{i\beta_2}(|0010\rangle+e^{i\beta_3}|0001\rangle)$. Hence, we can define $\Delta=1-F_1^2- \sum_{i=2}^{4} F_i^{2}$ in terms of the respective overlaps $F_i$. Because of the orthonormality $\sum_{i=1}^{4} F_i =1$, the sum uncertainty is bounded by $\Delta\geq 1-F_{1}^{2}-(1-F_{1})^{2}$, whereby we obtain $F_1\geq\sqrt{\frac{1}{2}(\frac{1}{2}-\Delta)}+\frac{1}{2}$. Finally, by combining the probability $\tilde p_1$ for exciting one spin-wave distributed among the four ensembles, we access the lower bound fidelity $\tilde{F}^{(A)}\geq\tilde{p}_{1}(\sqrt{\frac{1}{2}(\frac{1}{2}-\Delta)}+\frac{1}{2})$  for the heralded atomic state $\hat{\rho}^{(A)}_{W}$. In principle, the imbalances in the interferometer rotate the projectors into non-orthonormal sets\cite{lougovski09}. However, the measured losses and the beam-splitter ratios are sufficiently balanced such that any changes in $\tilde{F}^{(A)}$ due to modified projectors are well within the uncertainties of the data.

\noindent\textbf{Prospects for improving memory time and matter-light transfer efficiency.}
By operating the clock speed at $R_c\rightarrow10$ MHz and $\tau_m^{(3)} \approx 20 \mu s$, we can prepare hexapartite ($M=6$) entanglement with probability of $3z\eta_{\text{read}} p_h^2/8 \approx 10^{-5}$ by connecting two quadripartite states $\hat{\rho}^{(A)}_{W}$, where the enhancement factor $z$ is $400$ (ref. \cite{chou07}). However, the memory times $\{\tau^{(3)}_{m},\tau^{(2)}_{m},\tau^{(1)}_{m}\}$ in Fig. \ref{fig4} and the entanglement transfer $\lambda$ from matter to light limit our capability to scale the multipartite entanglement beyond $M>6$ by way of quantum control and swapping of entanglement \cite{chou07,laurat07a} with our current experimental parameters. The prerequisite storage techniques for suppressing both the internal and motional spin-wave dephasings can be extended for $\tau_{m}$ with advances in ensemble-based quantum memories\cite{hammerer10,rzhao08, bzhao08}. Recent experiments with single ensembles have achieved coherence times up to $\tau_m\simeq 1.5$ seconds in quantum degenerate gases \cite{schnorrberger09,zhang09} albeit with efficiencies $\lesssim 1\%$. The transfer efficiency can also be increased to $\lambda_{\text{th}}\simeq 0.9$ by enclosing the ensembles with high finesse cavities \cite{simon07b}. System integrations by way of atom-chip technology and waveguide coupling \cite{colombe07,vetsch10} hold great potential for scalability given the strong cooperativity and the long coherence \cite{deutsch10}. In addition, our experiment opens the door for future theoretical studies of complex repeater architectures for multipartite systems, beyond traditional one-to-one networks.

\vspace{0.4 cm}
\indent\textbf{Acknowledgement.} We gratefully acknowledge fruitful discussions with K. Hammerer, P. Zoller, and J. Ye.  This research is supported by the National Science Foundation, the DoD NSSEFF program, the Northrop Grumman Corporation, and the Intelligence Advanced Research Projects Activity. AG acknowledges support by the Nakajima Foundation. SP acknowledges support as a fellow of the Center for Physics of Information at Caltech. 
\end{document}